\def\be{\begin{equation}}
\def\ee{\end{equation}}
\def\ba{\begin{array}}
\def\ea{\end{array}}

\def\L{\Lambda}

\def\Nb{{I\!\! N}}

\documentstyle[12pt]{article}
\begin{document}
\parskip=4pt
\parindent=18pt
\baselineskip=22pt
\setcounter{page}{1}
\centerline{\Large\bf Some New Exact Ground States for Generalized}
\vspace{2ex}
\centerline{\Large\bf Hubbard Models}
\vspace{6ex}
\begin{center}
{\large  Sergio Albeverio}\footnote {SFB  237; BiBoS; CERFIM (Locarno); 
Acc.Arch., USI (Mendrisio)}~~~ and ~~~
{\large  Shao-Ming Fei}\footnote{Institute of Physics,
Chinese Academy of Science, Beijing.}

\end{center}
\begin{center}
Institut f\"ur Angew. Mathematik, Universit\"at Bonn\\
and\\
 Fakult\"at f\"ur Mathematik, Ruhr-Universit\"at Bochum\\
 D-44780 Bochum, Germany
\end{center}
\vskip 1 true cm
\parindent=18pt
\parskip=6pt
\begin{center}
\begin{minipage}{5in}
\vspace{3ex}
\centerline{\large Abstract}
\vspace{4ex}
A set of new exact ground states of the generalized Hubbard models
in arbitrary dimensions with explicitly given parameter regions is
presented. This is based on a simple method for constructing exact
ground states for homogeneous quantum systems.
\bigskip
\medskip

PACS numbers: 75.10.Jm, 71.27+a

\end{minipage}
\end{center}

\newpage
There has been a growing interest in constructing 
exact ground states for correlated Fermion systems in certain
parameter regions [1-7]. For one dimensional quantum systems, an
effective method to construct ground states is the so called matrix
product approach \cite{ksz}. Its origin can be traced back to the $S=1$
spin chain model \cite{chain}. For higher dimensional correlated
Fermion systems, there are also some approaches for the construction of
exact ground states \cite{hd}. In \cite{bs} the conditions for
$\eta$-pairing, paramagnetic, N\'eel,
charge-density-wave and ferromagnetic states to be exact ground states
of the generalized Hubbard model have been presented by using
the so called optimal ground state approach.
These approaches recovered and improved the
results obtained previously.

In this paper we give some new exact ground states for the
generalized Hubbard models. We first describe a general way
to construct some exact ground states
for homogeneous quantum systems. Let $H$ be the Hamiltonian of a quantum
lattice system (correlated Fermion system) of the form
$H=\sum_{<ij>}h_{ij}$, where $<ij>$ denotes the nearest
neighbours, $i,j$ are points on a finite lattice $\Gamma$ with $\L$ lattice 
sites, $H$ acts in the tensor space ${\cal H}$ associated with $\Gamma$ (i.e.
${\cal H}$ is the $\L$-fold power of some given Hilbert space $K$ associated
with a single lattice point), $h_{ij}$ is a self-adjoint operator acting on 
the tensor product $K\otimes K$ associated with $i,j$. We call
the system homogeneous if the spectral decomposition of $h_{ij}$ is 
independent of  $i,j$. For square lattice case, this homogeneity implies that
the horizontal and vertical couplings are the same.
The exact ground states of $H$ are by definition the eigenstates
belonging to the lowest eigenvalue of $H$. Let $\vert\xi_\alpha>$,
$\alpha=1,...,n$ be $n$ possible states of a local site, for instance,
empty, spin up (down) and double occupied states. We have, for some given
indices $\alpha$, $\beta\in 1,...,n$, the following conclusions:

I) If $\vert\xi_\alpha>_i\vert\xi_\alpha>_j$ is
an eigenvector of $h_{ij}$ and the corresponding eigenvalue is
the lowest one, the following state is
a ground state of the Hamiltonian $H$ of the system,
\be\label{i}
\psi^{I}=\prod_{i=1}^\L\vert\xi_\alpha>_i.
\ee

II) If $\vert\xi_\alpha>_i\vert\xi_\alpha>_j$,
$\vert\xi_\beta>_i\vert\xi_\beta>_j$ and
$\vert\xi_\alpha>_i\vert\xi_\beta>_j+\vert\xi_\beta>_i\vert\xi_\alpha>_j$
are eigenvectors of $h_{ij}$ with the same and
lowest energy, the state
\be\label{ii}
\psi^{II}=\prod_{i=1}^\L(\vert\xi_\alpha>_i+\vert\xi_\beta>_i)
\ee
is a ground state of $H$.

III) If $\vert\xi_\alpha>_i\vert\xi_\alpha>_j$ and
$\vert\xi_\alpha>_i\vert\xi_\beta>_j-\vert\xi_\beta>_i\vert\xi_\alpha>_j$
are eigenvectors of $h_{ij}$ with the same and lowest energy, the state
\be\label{iii}
\psi^{III}=\sum_{i=1}^\L(-1)^{i+1}\vert\xi_\alpha>_1\vert\xi_\alpha>_2
...\vert\xi_\beta>_i...\vert\xi_\alpha>_\L
\ee
is a ground state of $H$.

IV) If $\vert\xi_\alpha>_i\vert\xi_\alpha>_j$ and
$\vert\xi_\alpha>_i\vert\xi_\beta>_j+\vert\xi_\beta>_i\vert\xi_\alpha>_j$
are eigenvectors of $h_{ij}$ with the same and lowest energy, the state
\be\label{vi}
\psi^{IV}=\sum_{i=1}^\L\vert\xi_\alpha>_1\vert\xi_\alpha>_2
...\vert\xi_\beta>_i...\vert\xi_\alpha>_\L
\ee
is a ground state of $H$.

{\sf [Proof].} I is a consequence of homogeneity, II holds by a
simple calculation. As for III (IV can be similarly proved) we proceed
as follows. Let
$$
A_i=\left(\ba{ll}
\vert\xi_\alpha>_i&0\\[3mm]
\vert\xi_\alpha>_i+\vert\xi_\beta>_i&-\vert\xi_\alpha>_i
\ea
\right).
$$
We have
$$
A_iA_j=\left(\ba{ll}
\vert\xi_\alpha>_i\vert\xi_\alpha>_j&0\\[3mm]
\vert\xi_\alpha>_i\vert\xi_\beta>_j-\vert\xi_\beta>_i\vert\xi_\alpha>_j&
\vert\xi_\alpha>_i\vert\xi_\alpha>_j
\ea
\right).
$$
The nonzero elements $a_{ij}$ of $A_iA_j$ satisfy
$h_{ij}a_{ij}=\epsilon a_{ij}$, with $\epsilon$ the lowest eigenvalue
of $h_{ij}$. Therefore the elements of the matrix
$M\equiv\prod_{i=1}^\L A_i$, as well as their linear
combinations are ground states of $H$ with the same lowest eigenvalue
$P\epsilon$, where $P$ is the number of nearest-neighbour pairs of the
lattice. The linear combination of the elements $M_{(1,1)}$ and $M_{(2,1)}$
gives rise to the ground state $\psi^{III}$.
\hfill\rule{2mm}{2mm}

The state $\psi^{II}$ is a sum of terms containing factors
of the form $\vert\xi_\alpha>_i$ and $\vert\xi_\beta>_j$.
If there exists a local
operator $o$ such that $o\vert\xi_\alpha>=a_1\vert\xi_\alpha>$,
$o\vert\xi_\beta>=a_2\vert\xi_\beta>$, $a_1\neq a_2$ and the operator
$O=\sum_{i}^{\Lambda}o_{i}$ commutes with
the Hamiltonian $H$, then
\be\label{psil}
\psi^{II}_l=\sum_{m_1,...,m_l=1}^\L
\vert\xi_\alpha>_1...\vert\xi_\beta>_{m_1}...
\vert\xi_\beta>_{m_l}...\vert\xi_\alpha>_\L
\ee
are also ground states for all $l=0,1,...,\L$. In this case the states
$\psi^{II}_{l=1,\L-1}$ (resp. $\psi^{II}_{l=0,\L}$) are just the states
$\psi^{IV}$ (resp. $\psi^{I}$).
However, the state $\psi^{IV}$ (resp. $\psi^{I}$) has a
larger parameter region than $\psi^{II}$. In addition, all the states
$\psi^{II}_l$ are degenerate.

We now consider the generalized Hubbard model on an arbitrary
$d$-dimensional lattice with Hamiltonian \cite{bks},
\be\label{h}
\ba{rcl}
H&=&\sum_{<ij>}^\Lambda h_{ij}\equiv
\sum_{<ij>}^\Lambda\left[-t
\sum_{\sigma}(c_{i\sigma}^{\dagger}c_{j\sigma}
+c_{j\sigma}^{\dagger}c_{i\sigma})
+X\sum_{\sigma}(c_{i\sigma}^{\dagger}c_{j\sigma}
+c_{j\sigma}^{\dagger}c_{i\sigma})(n_{i,-\sigma}+n_{j,-
\sigma})\right.\\[5mm]
&&+
\frac{U}{Z}\left(\left(n_{i\uparrow}-\frac{1}{2}\right)
\left(n_{i\downarrow}-\frac{1}{2}\right)+
\left(n_{j\uparrow}-\frac{1}{2}\right)
\left(n_{j\downarrow}-\frac{1}{2}\right)\right)
+V(n_i-1)(n_j-1)\\[4mm]
&&\left.
+Y(c_{i\uparrow}^{\dagger}c_{i\downarrow}^{\dagger}
c_{j\downarrow}c_{j\uparrow}
+c_{j\uparrow}^{\dagger}c_{j\downarrow}^{\dagger}
c_{i\downarrow}c_{i\uparrow})
+\frac{J_{xy}}{2}(S_i^+ S_j^- + S_j^+ S_i^-) +J_z S_i^z S_j^z 
+\frac{\mu}{Z}(n_i+n_j)\right],
\ea
\ee
where $c_{i\sigma}^{\dagger}$ and $c_{i\sigma}$, 
$\sigma=\uparrow,\downarrow$, are canonical Fermi creation resp.
annihilation
operators, $n_{i,\sigma}=c_{i\sigma}^{\dagger}c_{i\sigma}$ and
$n_i=n_{i\uparrow}+n_{i\downarrow}$ are the number operators,
$S_i^+=c_{i\uparrow}^{\dagger}c_{i\downarrow}$,
$S_i^-=c_{i\downarrow}^{\dagger}c_{i\uparrow}$,
$S_i^z=(n_{i\uparrow}-n_{i\downarrow})/2$; $\mu$ is the chemical
potential and $Z$ the coordination number of the $d$-dimensional
lattice; $t$, $X$, $U$, $V$, $Y$, $J_{xy}$ and $J_z$ are real valued
coupling constants for, respectively,
single particle hopping, bond-charge interaction,
on-site Coulomb interaction, nearest-neighbour Coulomb interaction,
pair-hopping and XXZ-type spin interactions.

The well known states related to the Hubbard model are the 
$\eta$-pairing, paramagnetic, N\'eel, charge-density-wave (CDW)
and ferromagnetic (F) states, defined respectively as follows:
\be\label{ws}
\ba{l}
\vert N{\acute e}el>=\prod_{i\in B}c_{i\uparrow}^{\dagger}
\prod_{i\in B^\prime}c_{i\downarrow}^{\dagger}\vert 0>,~~~
\vert\psi_m^\eta>=\left(\sum_{j=1}^\Lambda e^{i\eta j}
c_{j\downarrow}^{\dagger}
c_{j\uparrow}^{\dagger}\right)^m\vert 0>,~~~m\in\Nb\\[5mm]
\vert para>=\prod_{i\in A}c_{i\uparrow}^{\dagger}
\prod_{i\in A^\prime}c_{i\downarrow}^{\dagger}\vert 0>,~~~
\vert CDW>=\prod_{i\in B}c_{i\uparrow}^{\dagger}
c_{i\downarrow}^{\dagger}\vert 0>,~~~
\vert F>=\prod_{i}c_{i\uparrow}^{\dagger}\vert 0>,
\ea
\ee
where $\vert 0>$ is the vacuum state, 
$A\cap A^\prime=\emptyset$, $A$ and $A^\prime$ together 
span the whole lattice,
$B$ and $B^\prime$ are the odd and even sublattices of a bipartite
lattice, $\eta$ is a real parameter.
The $\eta$-pairing states display off diagonal long-range order (ODLRO)
\cite{ya}, i.e. superconductivity (Meissner effect and flux quantization 
\cite{yz}).

The exact ground states of $H$ depend on the coupling constants.
At certain parameter regions the above states (\ref{ws})
become exact ground states \cite{hd,bs},
as also follows from our conclusions. We observe that
some of the ground states could be physically equivalent owing to
some symmetries of the system in certain parameter regions, for
instance, when $J_z=J_{xy}$, $H$ is $SU(2)$ invariant.
Nevertheless besides the well known exact ground states the
quantum system (\ref{h}) has many additional exact ground states of
types $\psi^{II}$, $\psi^{III}$ and $\psi^{IV}$ with 
respect to different parameter regions. In the
following we simply list some of these new ground states.

We first present some exact ground eigenstates of the form $\psi^{III}$.
\be\label{ps1}
\vert\psi_1^\sigma>=N\sum_{i=1}^\Lambda (-1)^{i+1}c_{1\sigma}^{\dagger}
c_{2\sigma}^{\dagger}...c_{i,-
\sigma}^{\dagger}...c_{\Lambda\sigma}^{\dagger}\vert 0>,~~~
N\equiv(\frac{1}{2})^{\frac{\L}{2}}
\ee
is a ground state when $t=X$, $J_z=-J_{xy}<0$,
$\frac{U}{Z}\geq max(\frac{J_z}{4}
+\frac{2\mu}{Z}-V,~\frac{J_z}{2}+\frac{2\mu}{Z}+2\vert
t\vert,~\frac{J_z}{4}+V+Y,~
\frac{J_z}{2}-\frac{2\mu}{Z}+2\vert t\vert,~\frac{J_z}{4}-
\frac{2\mu}{Z}-V,~\frac{J_z}{4}+V-Y)$.
The eigenvalue is $(-\frac{U}{2Z}+\frac{J_z}{4}+\frac{2\mu}{Z})P$. If one
sets $J_z=J_{xy}=0$ instead of $J_z=-J_{xy}<0$, state
(\ref{ps1}) is still a ground state but is then equivalent to the 
$\vert para>$ state, because $\vert\psi_1^\sigma>$ is a linear combination
of $\vert para>$ states, while under condition $t=X$, $J_z=J_{xy}=0$, any
$\vert para>$ state is an exact ground state.

The state
\be\label{ps2}
\vert\psi_2^\sigma>=\sum_{i=1}^\Lambda (-1)^{i+1}c_{1\sigma}^{\dagger}
c_{2\sigma}^{\dagger}...c_{i,\uparrow}^{\dagger}c_{i,\downarrow}^{\dagger}
...c_{\Lambda\sigma}^{\dagger}\vert 0>
\ee
is a ground state at parameter region
$X+\frac{\mu}{Z}\leq 0$ if $t\geq 0$,
$X+\frac{\mu}{Z}\leq t$ if $t\leq 0$,
$J_z=\frac{2U}{Z}+\frac{4\mu}{Z}-4(t-2X)$
and
$max (-t+2X+\frac{3\mu}{Z}-V,
~-t+2X+V+Y+\frac{\mu}{Z},~-t+2X-
V-\frac{\mu}{Z})\leq \frac{U}{2Z}
<V-Y+\frac{\mu}{Z}-t+2X$,
$\frac{J_{xy}}{4}\geq max(\frac{U}{2Z}-t+2X+\frac{\mu}{Z},~
t-2X-\frac{\mu}{Z}-\frac{U}{2Z}+2(X-t)^2/(V-Y-t+2X+\frac{\mu}{Z}-
\frac{U}{2Z}))$, or
$\frac{U}{2Z}\geq 
max(-t+2X+\frac{3\mu}{Z}-V,~-
t+2X+V+Y+\frac{\mu}{Z},~-t+2X-
V-\frac{\mu}{Z})$, $\frac{U}{2Z}>V-Y+\frac{\mu}{Z}-t+2X$,
$t-2X-\frac{\mu}{Z}-\frac{U}{2Z}+2(X-t)^2/(V-Y-t+2X
+\frac{\mu}{Z}-\frac{U}{2Z}) \geq\frac{J_{xy}}{4}\geq 
\frac{U}{2Z}-t+2X+\frac{\mu}{Z}$
for $t\geq 2X$, $t\neq X$, or $max (t-V+\frac{3\mu}{Z},
~t+\frac{\mu}{Z}+V+Y,~t-V-\frac{\mu}{Z},~t+V+\frac{\mu}{Z}-Y
)\leq \frac{U}{2Z}
\leq min(\frac{J_{xy}}{4}-t-\frac{\mu}{Z},~
-\frac{J_{xy}}{4}-t-\frac{\mu}{Z})$,
$J_z=\frac{2U}{Z}+\frac{4\mu}{Z}+4t$, $\mu\leq 0$
for $t=X\leq 0$. The eigenvalue of $\vert\psi_2^\sigma>$ is $(2X-
t+\frac{3\mu}{Z})P$.

For $t\leq 0$ and $J_z=4(t+\frac{U}{2Z}-\frac{\mu}{Z})$ the following
state 
\be\label{ps4}
\vert\psi_3^\sigma>=N\sum_{i=1}^\Lambda (-1)^{i+1}
c_{1,\sigma}^{\dagger}c_{2,\sigma}^{\dagger}...
{\bf 1}_i...c_{\Lambda,\sigma}^{\dagger}\vert 0>,
\ee
${\bf 1}$ being the identity operator,
is an exact ground state, for $t\neq X$, 
$\frac{\mu}{Z}\geq X\geq t-\frac{\mu}{Z}$ and
$V-Y+t-\frac{\mu}{Z}>\frac{U}{2Z}
\geq max(t+\frac{\mu}{Z}-V,
~t-\frac{\mu}{Z}+V+Y,~t-V-\frac{3\mu}{Z})$,
$\frac{J_{xy}}{4}\geq max(\frac{U}{2Z}
+t-\frac{\mu}{Z},~\frac{\mu}{Z}-t-
\frac{U}{2Z}+2(X-t)^2/(V-Y+t-\frac{U}{2Z}-
\frac{\mu}{Z}))$ or
$\frac{U}{2Z}\geq max(t+\frac{\mu}{Z}-V,
~t-\frac{\mu}{Z}+V+Y,~t-V-\frac{3\mu}{Z}
)$, $\frac{U}{2Z}>V-Y+t-\frac{\mu}{Z}$,
$\frac{\mu}{Z}-t-\frac{U}{2Z}+2(X-t)^2/(V-Y+t-\frac{U}{2Z}-
\frac{\mu}{Z})\geq\frac{J_{xy}}{4}\geq\frac{U}{2Z}+t-\frac{\mu}{Z}$,
or for $t=X\leq 0$, $\mu\geq 0$ and
$max (t+\frac{\mu}{Z}-V,
~t-\frac{\mu}{Z}+V+Y,~t-\frac{\mu}{Z}+V-Y
)\leq \frac{U}{2Z}\leq min(
-t+\frac{\mu}{Z}+\frac{J_{xy}}{4},~-t-
\frac{J_{xy}}{4}+\frac{\mu}{Z})$.

Another kind of exact ground states of the generalized Hubbard model of
the form $\psi^{IV}$ is given by:
\be\label{ps6}
\vert\psi_4^\sigma>=\sum_{i=1}^\Lambda c_{1\sigma}^{\dagger}
c_{2\sigma}^{\dagger}...c_{i,\uparrow}^{\dagger}c_{i,\downarrow}^{\dagger}
...c_{\Lambda\sigma}^{\dagger}\vert 0>.
\ee
The corresponding parameter region is, for $t\leq 2X$, $t\neq X$,
$J_z=4(\frac{U}{2Z}+\frac{\mu}{Z}+t-
2X)$, $X-\frac{\mu}{Z}\geq t$ if
$t\geq 0$, $X-\frac{\mu}{Z}\geq 0$ if $t\leq 0$,
$\frac{U}{2Z}\geq max(t-2X-
V+\frac{3\mu}{Z},~t-2X+V+Y+\frac{\mu}{Z},~t-
2X-\frac{\mu}{Z}-V)$ and
$\frac{U}{2Z}<V-Y+t-2X+\frac{\mu}{Z}$,
$\frac{J_{xy}}{2}\geq max(\frac{U}{Z}+2t-
4X+\frac{2\mu}{Z},~4X-2t-\frac{U}{Z}-
\frac{2\mu}{Z}+4(X-t)^2/(V-Y+t-2X-
\frac{U}{2Z}+\frac{\mu}{Z}))$ or
$\frac{U}{2Z}>V-Y+t-2X+\frac{\mu}{Z}$,
$4X-2t-\frac{U}{Z}-
\frac{2\mu}{Z}+4(X-t)^2/(V-Y+t-2X-
\frac{U}{2Z}+\frac{\mu}{Z})\geq
\frac{J_{xy}}{2}\geq\frac{U}{Z}+2t-
4X+\frac{2\mu}{Z}$, and for $t=X\geq 0$,
$J_z=4(\frac{U}{2Z}+\frac{\mu}{Z}-t)$, $\mu\leq 0$,
$min(\frac{J_{xy}}{4}+t-\frac{\mu}{Z},
t-\frac{J_{xy}}{4}-\frac{\mu}{Z})\geq
\frac{U}{2Z}\geq max(-t+V+Y+\frac{\mu}{Z},
-t-\frac{\mu}{Z}-V,~-t+V-Y+\frac{\mu}{Z})$.
$\vert\psi_6^\sigma>$ corresponds to the eigenvalue $(t-2X+\frac{3\mu}{Z})P$.

For $t\geq 0$ and $J_z=4(-t+\frac{U}{2Z}-\frac{\mu}{Z})$ the state 
\be\label{ps8}
\vert\psi_5^\sigma>=N\sum_{i=1}^\Lambda
c_{1,\sigma}^{\dagger}c_{2,\sigma}^{\dagger}...
{\bf 1}_i...c_{\Lambda,\sigma}^{\dagger}\vert 0>
\ee
is an exact ground state, for $t\neq X$, 
$-\frac{\mu}{Z}\leq X\leq \frac{\mu}{Z}+t$,
and $V-Y-t-\frac{\mu}{Z}>\frac{U}{2Z}
\geq max(-t+\frac{\mu}{Z}-V,
~-t-\frac{\mu}{Z}+V+Y,~-t-V-\frac{3\mu}{Z})$,
$\frac{J_{xy}}{4}\geq max(\frac{U}{2Z}
-t-\frac{\mu}{Z},~\frac{\mu}{Z}+t-
\frac{U}{2Z}+2(X-t)^2/(V-Y-t-\frac{U}{2Z}-\frac{\mu}{Z}))$
or $\frac{U}{2Z}
\geq max(-t+\frac{\mu}{Z}-V,
~-t-\frac{\mu}{Z}+V+Y,~-t-V-\frac{3\mu}{Z}
)$, $\frac{U}{2Z}>V-Y-t-\frac{\mu}{Z}$,
$\frac{\mu}{Z}+t-
\frac{U}{2Z}+2(X-t)^2/(V-Y-t-\frac{U}{2Z}-\frac{\mu}{Z})\geq
\frac{J_{xy}}{4}\geq\frac{U}{2Z}
-t-\frac{\mu}{Z}$, or for $t=X$, $\mu\geq 0$, and
$max (-t+\frac{\mu}{Z}-V,
~-t-\frac{\mu}{Z}+V+Y,~-t-\frac{\mu}{Z}+V-Y
)\leq \frac{U}{2Z}\leq min(
t+\frac{\mu}{Z}+\frac{J_{xy}}{4},~t-
\frac{J_{xy}}{4}+\frac{\mu}{Z}$.
The eigenvalue is $(-t+\frac{\mu}{Z})P$.

The states of the form $\psi^{II}$ are sums of states with different
number of electrons. As the Hamiltonian commutes with the
number operator, any projection of a ground state of the type $\psi^{II}$
to a state with a given number of electrons is also a ground state.
These ground states are of the form (\ref{psil}), for instance,
for $l=2,...,\L-2$, the following states are ground states,
\be\label{ps10}
\vert\psi_{l}^\sigma>=\sum_{m_1,...,m_l=1}^\L
c_{1\uparrow}^{\dagger}c_{1\downarrow}^{\dagger}...
c_{m_1\sigma}^{\dagger}...c_{m_l\sigma}^{\dagger}...
c_{\L\uparrow}^{\dagger}c_{\L\downarrow}^{\dagger}\vert 0>
\ee
if the parameters satisfy 
$J_z=4(\frac{U}{2Z}-t+\frac{\mu}{Z})$,
$V=-\frac{\mu}{Z}-t-\frac{U}{2Z}$,
$\mu\leq 0$, $\vert Y\vert-2t\leq\frac{U}{Z}\leq 2t-
\frac{\vert J_{xy}\vert}{2}-\frac{2\mu}{Z}$
for $t=X>0$. For $t\neq X$ we have
$J_z=4(\frac{U}{2Z}+t-2X+\frac{\mu}{Z})$,
$V=-\frac{\mu}{Z}+t-2X-\frac{U}{2Z}$,
$\mu\leq 0$, $X\geq max
(\frac{\mu}{Z},t+\frac{\mu}{Z}
,\frac{t}{2})$ and
$2t-4X+Y\leq\frac{U}{Z}<2t-4X-Y$,
$\frac{J_{xy}}{2}
\geq max(\frac{U}{Z}+2t-4X+\frac{2\mu}{Z},~
4X-2t-\frac{2\mu}{Z}-\frac{U}{Z}
+4(X-t)^2/(2t-4X-Y-\frac{U}{Z}))$ or
$2t-4X+Y\leq\frac{U}{Z}$, $2t-4X-Y<\frac{U}{Z}$,
$4X-2t-\frac{2\mu}{Z}-\frac{U}{Z}
+4(X-t)^2/(2t-4X-Y-\frac{U}{Z})
\geq\frac{J_{xy}}{2}
\geq \frac{U}{Z}+2t-4X+\frac{2\mu}{Z}$.
The eigenvalues are $(\frac{3\mu}{Z}+t-2X)P$.
For $l=0,1,\L-1,\L$, the states $\vert\psi_{l}^\sigma>$ are already
included in $\psi^I$ and $\psi^{IV}$.

Similarly we have the following ground states:
\be\label{ps11}
\vert\phi_{l}^\sigma>=\sum_{m_1,...,m_l=1}^\L
c_{m_1\sigma}^{\dagger}...c_{m_l\sigma}^{\dagger}\vert 0>,~~~l=2,...,\L-2
\ee
when
\be\label{31}
V=\frac{\mu}{Z}-t-\frac{U}{2Z},~~~~
J_z=4(\frac{U}{2Z}-t-\frac{\mu}{Z})
\ee
and, for $t\neq X$, $t\geq 0$,
$Y-2t\leq\frac{U}{Z}<-Y-2t$, $\mu\geq 0$,
$\frac{\mu}{Z}+t\geq X\geq -\frac{\mu}{Z}$,
$\frac{J_{xy}}{4}\geq max(
-t-\frac{\mu}{Z}+\frac{U}{2Z}
,~-\frac{U}{2Z}+t
+\frac{\mu}{Z}-2(X-t)^2/(\frac{U}{Z}+Y+2t))$
or $Y-2t\leq\frac{U}{Z}$,
$-Y-2t<\frac{U}{Z}$, $\mu\geq 0$,
$\frac{\mu}{Z}+t\geq X\geq -\frac{\mu}{Z}$,
$-\frac{U}{2Z}+t
+\frac{\mu}{Z}-2(X-t)^2/(\frac{U}{Z}+Y+2t)\geq
\frac{J_{xy}}{4}\geq
-t-\frac{\mu}{Z}+\frac{U}{2Z}$,
or for $t=X\geq 0$,
\be\label{322}
-2t+\vert Y\vert\leq\frac{U}{Z}\leq 2t-\frac{\vert
J_{xy}\vert}{2}+\frac{2\mu}{Z},~~~~\mu\geq 0.
\ee
The corresponding eigenvalue is $(\frac{\mu}{Z}-t)P$.

From formulae (1-4) one can construct more exact ground states such as
$\vert\psi_6^\sigma>=N\sum_{i=1}^\Lambda (-1)^{i+1}
c_{1,\uparrow}^{\dagger}c_{1,\downarrow}^{\dagger}
c_{2,\uparrow}^{\dagger}c_{2,\downarrow}^{\dagger}...
c_{i\sigma}^{\dagger}...
c_{\Lambda,\uparrow}^{\dagger}c_{\Lambda,\downarrow}^{\dagger}\vert 0>$,
$\vert\psi_7^\sigma>=N\sum_{i=1}^\Lambda (-1)^{i+1}
c_{i,\sigma}^{\dagger}\vert 0>$,
$\vert\psi_8^\sigma>=N\sum_{i=1}^\Lambda
c_{1,\uparrow}^{\dagger}c_{1,\downarrow}^{\dagger}
c_{2,\uparrow}^{\dagger}c_{2,\downarrow}^{\dagger}...
c_{i\sigma}^{\dagger}...
c_{\Lambda,\uparrow}^{\dagger}c_{\Lambda,\downarrow}^{\dagger}\vert 0>$,
$\vert\psi_{9}^\sigma>=N\sum_{i=1}^\Lambda
c_{i,\sigma}^{\dagger}\vert 0>$ in certain parameter regions. 
Some of the states have the same eigenvalues. For instance,
the state $\vert\psi_6^\sigma>$ has the same eigenvalue as
$\vert\psi_2^\sigma>$, however
$\vert\psi_2^\sigma>$ and $\vert\psi_6^\sigma>$
correspond to different parameter regions. Thus in general
the eigenvalues are not degenerate.

The $h_{ij}$ in (\ref{h}) have two different sets of eigenvectors and
eigenvalues for $t=X$ and $t\neq X$ respectively. Therefore the parameter
regions of the ground states have a jump when $t\to X$. In addition, all the
ground states we presented above have no ODLRO. A ground state with
ODLRO can be obtained from formula (\ref{psil}) by taking
$\vert\xi_\alpha>=\vert 0>$ and
$\vert\xi_\beta>=c_{\uparrow}^{\dagger}c_{\downarrow}^{\dagger}\vert 0>$.
One then gets ground states that are equivalent to the $\eta$-pairing
states with $\eta=0$:
$\vert\psi_{l}^{\eta=0}>=\sum_{m_1,...,m_l=1}^\L
c_{m_1\uparrow}^{\dagger}c_{m_1\downarrow}^{\dagger}...
c_{m_l\uparrow}^{\dagger}c_{m_l\downarrow}^{\dagger}\vert 0>$
within the parameter region 
$\frac{U}{Z}\leq min(-2\vert t\vert-2V,\-~\frac{J_z}{4}-V,\-~
-V-\frac{\vert J_{xy}\vert}{2}-\frac{J_z}{4})$,
$t=X$, $Y=2V$, $\mu=0$ and $V\leq 0$.

Some properties of the phase
diagram of the ground states can be analysed according to their related
parameter regions. For instance, taking into account the conditions
(\ref{31}) and (\ref{322}) (for simplicity we set $J_{xy}=0$), we have
that $\vert\psi_{l}^{\eta=0}>$ is a ground state for
$\frac{U}{Z}\leq\vert 2t\vert$. When $\frac{U}{Z}<-2t$, the
$\eta$-pairing $\vert\psi_{l}^{\eta=0}>$
is no longer a ground state, but $\vert\phi_{2l}^\sigma>$
given by (\ref{ps11}) is a ground state.

Some relations between the ground states above and the ground states
studied in \cite{bs} can be discussed according to their parameter
regions. As an example we set
$t=X$, $Y=\mu=0$, $V\geq 0$, and $J_z=-J_{xy}<0$.
It can be shown that for
$\vert t\vert\geq J_{xy}/8$ the states
$\vert\psi_1^\sigma>$ and $\vert F>$ are ground states
in regions I and IV. While the state $\vert CDW>$
is a ground state in region II and III, see Fig. 1.
More properties can be obtained when one changes the relations among
the parameters. For instance, if $t=V$, then
the states $\vert\psi_1^\sigma>$ and $\vert F>$ (resp. $\vert CDW>$)
are ground states only in region I (resp. region II).

\begin{center}
\begin{picture}(300,150)(-20,0)
\put(0,0){\line(1,1){150}}
\put(75,75){\line(1,2){37}}
\put(0,0){\line(2,0){225}}
\put(0,150){\line(2,0){225}}
\put(75,75){\line(2,0){150}}
\put(0,0){\line(0,2){150}}
\put(225,0){\line(0,2){150}}
\put(-10,140){$\frac{U}{Z}$}
\put(38,138){$\frac{U}{Z}=2V+\frac{J_{z}}{2}$}
\put(150,138){$\frac{U}{Z}=V+\frac{J_{z}}{4}$}
\put(75,65){\footnotesize ($0,-\frac{J_z}{4}$)}
\put(12,4){\footnotesize ($\frac{J_z}{4},0$)}
\put(75,75){\circle*{2}}
\put(0,0){\circle*{2}}
\put(-10,72){0}
\put(230,72){V}
\put(25,110){$\vert\psi_1^\sigma>$}
\put(25,95){$\vert F>$}
\put(165,30){$\vert CDW>$}
\put(150,95){III}
\put(110,122){IV}
\put(110,30){II}
\put(25,65){I}
\end{picture}
\end{center}

\begin{figure}[ht]
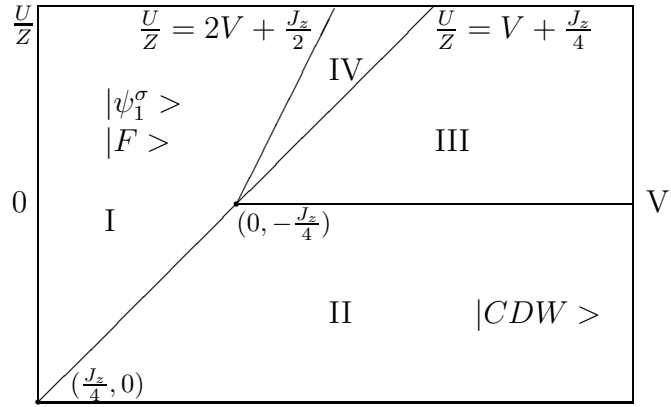

\caption
{A ground state phase diagram related to states $\vert F>$,
$\vert CDW>$ and $\vert\psi_1^\sigma>$}
\end{figure}

\end{document}